# CU Virginis – The First Stellar Pulsar

B. J. Kellett[1*], Vito G. Graffagnino[1], Robert Bingham[1,2], Tom W. B. Muxlow[3] & Alastair G. Gunn[3].

[1]*Rutherford Appleton Laboratory, Space Science & Technology Department, Chilton, Didcot OX11 0QX, UK.*

[2]*Dept. of Physics, University of Strathclyde, Glasgow, G4 0NG, U.K.*

[3]*MERLIN/VLBI National Facility, Jodrell Bank Observatory, The University of Manchester, Macclesfield, Cheshire SK11 9DL, UK.*

[*]*To whom correspondence should be addressed; E-mail: B.J.Kellett@rl.ac.uk..*

**CU Virginis is one of the brightest radio emitting members of the magnetic chemically peculiar (MCP) stars and also one of the fastest rotating. We have now discovered that CU Vir is unique among stellar radio sources in generating a persistent, highly collimated, beam of coherent, 100% polarised, radiation from one of its magnetic poles that sweeps across the Earth every time the star rotates. This makes the star strikingly similar to a pulsar. This similarity is further strengthened by the observation that the rotating period of the star is lengthening at a phenomenal rate (significantly faster than any other astrophysical source — including pulsars) due to a braking mechanism related to it's very strong magnetic field.**

CU Vir (HD124224, HR5313) was discovered as a spectrum variable star in 1952 (1) and as a stellar radio source in 1994 (2). Its rotation period of close to half a day was immediately recognised as was the fact that it had a strong magnetic field and that it rotated perpendicular to our line-of-sight (1). It is now known to be one of the brightest radio sources in the class of magnetic chemically peculiar (MCP) stars. It is detected from 1.4 to 88 GHz with fluxes between 1 and 5 mJy, peaking at around 15 GHz (3, 4). However, these stars are also known to possess strong dipolar magnetic fields and have their magnetic axis offset from the rotation axis of the star. This dipole axis tilt results in our line of sight to the star passing through different parts of the stellar magnetosphere as the star rotates. CU Vir is especially suitable for observations because it is relatively close (80 pc) and has a short rotation period of just 12.5 hours (0.52 days). However, perhaps its most useful physical characteristic is the previously referenced tilt of the magnetic axis of the star relative to its rotation axis. This tilt ensures that our line-of-sight to the star passes through the magnetic equator twice during each stellar rotation (5).

Pulsars were discovered nearly forty years ago yet, despite intensive study, the mechanism for generating their radio emission remains an unsolved puzzle (6). They are rapidly spinning neutron stars that produce highly regular pulses in a manner analogous to a lighthouse beam. Their spin periods increase with time at a generally steady rate which is due to the phenomenon of magnetic braking — the effect of spinning a magnetic field of $10^{12}$ G exerts a torque that slows the pulsar down (7). We know the radio emission from pulsars is highly nonthermal, covers a broad range of radio frequencies, originates from a very compact region with a super strong magnetic field and involves both electrons and positrons. However, all these factors also make pulsars extremely difficult to probe for details that might elucidate the emission mechanism. The extreme brightness temperature of pulsar radio emission requires a coherent generation process which undoubtedly involves a plasma instability of some sort. One interesting point worthy of note is that the "oblique rotator" model of rapidly rotating neutron stars was actually first developed to explain the periodic variations of magnetic field strength in stars like CU Vir (8) (i.e. more than twelve years before pulsars were first discovered).

In a series of recent papers (9–11), we have outlined a new electron cyclotron maser emission mechanism and it is now a laboratory experiment (12). This new mechanism is relevant to several types of astrophysical sources, including stellar radio sources. The basic requirements of the mechanism are a dipole magnetic field and a mildly energetic (i.e. keV) electron beam. The mechanism generates 100% right-hand circularly polarized emission. CU Vir is a star that perfectly epitomizes these requirements and could therefore be an ideal stellar laboratory for studying maser radiation generation. The first indications of maser radiation were revealed by the VLA in 1998 (13). The VLA data covered only a limited fraction of the rotation period of CU Vir but did detect two enhancements in the L-band flux (21cm/1.4 GHz) of the star that were both 100% right circularly polarised, exactly what our maser mechanism predicted. Here, we demonstrate for the first time that this phenomena is long lasting (persisting for at least eight years) and recurring (i.e. seen on four out of four rotations, periodically, again as our model predicts).

We observed CU Vir with the Multi-Element Radio-Linked Interferometer Network (MERLIN) at 1.658 GHz in May, 2006, on four consecutive nights (23rd–26th). The flux in the right and left-hand circular polarizations are shown for each individual night in figure 1(a) and for all four nights combined together in figure 1(b). The data are plotted here in rotation phase of the star – i.e. folded on the ~12.5 hour rotation period with an arbitrary phase zero of 0 UT on the first day (May 23$^{rd}$). An extra half cycle of phase is included in figure 1 to emphasize that the two emission peaks are not symmetrical in phase. It is these two emission peaks that are the most striking features of the data. Both peaks are essentially 100% right-hand polarized and appear to be roughly equal in average strength and width. Figure 1(b) indicates that the peaks are each about 13-14 degrees wide – or 28±1 minutes in time (FWHM). Any period close to 12.5 hours gives very similar results since the data shown here only cover seven rotations of the star (i.e. every alternate rotation for four consecutive nights). The peak at around phase 0.8 in figure 1 was the main target of our observing campaign (hence the fact that it was seen on all 4 nights). However, it was not seen at the exact time that we had predicted in advance based on the previously known rotation period of the star (14) and the previously known position of the enhancement (13). The peak was actually detected some 1.7 hours later than predicted – a significantly large error for something that should be well determined. However, CU Vir has exhibited significant period changes in the past (14) when the long term optical monitoring data for the star could be explained by two constant periods (12.496267 and 12.496874 hours). The longer period applies after around 1984. The period jump in 1984 amounts to 2.185 seconds. If we assume that the 1.7 hour delay in the time we observed the peak was due to a period jump of the same magnitude as in 1984 then the new period is 12.497481 hours and the second jump would have occurred around 4 years before the MERLIN observation (i.e. around May, 2002). This shows the value and diagnostic potential of repeated monitoring observations in the radio of these systems.

Our radio emission mechanism requires that the radiation is generated in a low density region above the magnetic pole (or poles) of a dipole magnetic geometry (9). Observations of CU Vir suggest that its magnetic field is offset from the centre of the star so that one magnetic pole is significantly stronger than the other (15). The radio observations of CU Vir clearly shows two strong enhancements which we would suggest originate from the two sides of a wide emission cone emerging from the stronger magnetic pole as opposed to one enhancement from each magnetic pole. If the emission originated at the two poles of the star it should be symmetric in phase.

Figure 1 clearly demonstrates that the two enhancements are far from symmetrically distributed. Figure 2 presents a simplified schematic picture of CU Vir – viewed from above the stellar rotation pole and assuming that the magnetic field is perpendicular to the rotation axis (the actual offset angle is around $70^o$ (15)). The radio emission from the stronger magnetic pole of the star will be emitted into a cone with a wide opening angle but a narrow thickness (the two extreme edges of the cone are shown in figure 2). The observations would then suggest that the cone opening angle is around $150^o$ and the cone thickness is around $14^o$. The inclination of the stellar rotation axis to our line-of-sight is estimated to be $60^o$ (15) which with the magnetic dipole angle of $70^o$ suggests our line-of-sight will pass very close to the magnetic poles of the star.

We suggest that CU Vir could be a "laboratory" for studying pulsar emission mechanisms. The radio properties of CU Vir do bear some striking similarities to pulsars. The radiation is clearly being beamed into space (although, admittedly, with a rather different emission geometry). This difference is due to CU Vir only generating keV energy electrons in its magnetosphere. In a pulsar, ultra-relativistic electrons (and positrons) are responsible for the emission and this results in the emission pattern being strongly beamed forward. Also, the presence of positrons in pulsars and not in CU Vir can explain the lack/presence of circular polarisation. In pulsars, the electrons would be say 100% right-hand polarised while the positrons would be the opposite, say 100% left-hand polarised. For roughly equal populations of electrons and positrons this will lead to no nett polarisation in the final combined beam. CU Vir only involves electrons, so it is 100% right circularly polarised. We have now demonstrated that the radio emission is persistent – it was seen on all four nights of our observation and we must assume it has been present at least since the VLA observations in 1998 (13). (Both CU Vir and radio pulsars emit all the time – the "pulsing" nature of the emission is determined by the magnetic axis of the source being offset from the rotation axis and the subsequent lighthouse pattern sweeping across our line-of-sight). These characteristics of CU Vir are unique among radio emitting stellar sources. CU Vir is therefore the first member of what could be termed "pulstars" — stellar pulsars. CU Vir clearly differs from pulsars in its size, rotation rate and the fact that so far the radio enhancements are confined to the L-band. The enhancements were not seen in the 5GHz/6 cm band (or higher frequencies) (13). However, these differences could be advantageous as the larger size of the emission region and the non-relativistic nature of the electrons involved makes probing CU Vir much easier than pulsars (plus CU Vir is much closer than the typical pulsar). A second distinctive characteristic of CU Vir that is similar to pulsars is its rapidly changing rotation period. CU Vir is virtually unique among stellar sources in showing clear observational evidence of its rotation period increasing and the magnitude of the change is also exceptional ($\approx$4 seconds in two "jumps" in ~twenty years). Another MCP star, 56 Ari, has also shown increases in its rotation period but only at the level of 2 seconds/century (16). Period changes have also been seen in contact binary systems, but these are caused by mass transfer events between the two stars in the binary (e.g. 44ι Bootis shows a long-term period increase of $2.7\times10^{-10}$ s/s (17)). Both CU Vir and 56 Ari are clearly identified as isolated, single, rapidly rotating, stars. CU Vir's spin-down rate is approximately $5\times10^{-9}$ s/s (or 20 seconds/century). This is a rate some 10,000 times faster than the Crab pulsar (for example). It is possible that the sudden changes in rotation period suggested by the observations could be due to structural changes in the outer envelope of the star caused by its very strong magnetic field (18). CU Vir is therefore giving us a glimpse of a phase of stellar evolution that all single stars must go through – the rapid spin-down from stellar birth to quiescent Main Sequence existence. This is an extremely rapid phase in most stars evolution and therefore the chance to observe the process "live" is likely to be highly informative. Long-term monitoring of CU Vir will be able to reveal more details of the maser emission mechanism and whether the process of stellar spin-down is episodic or pseudo-continuous.

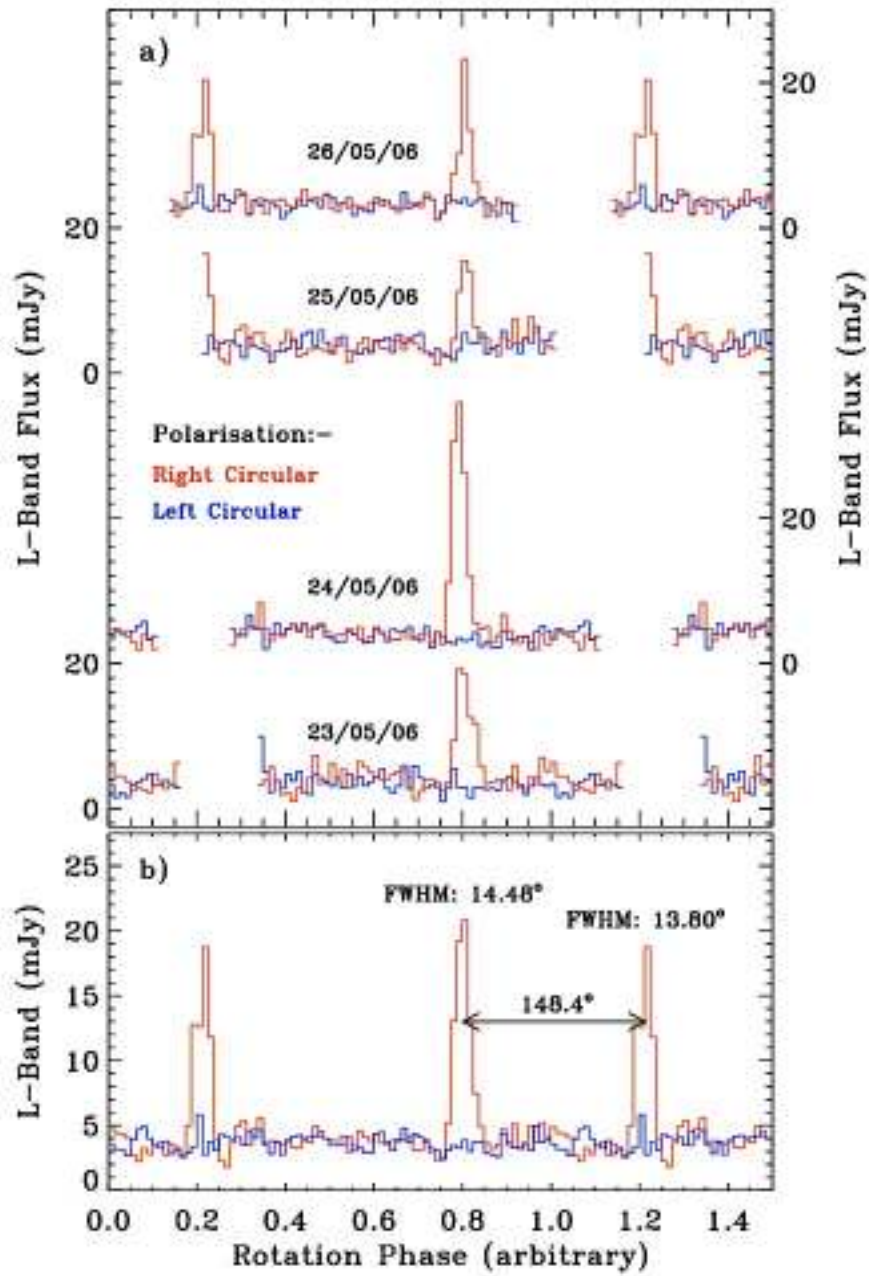

Figure 1: (a) The four individual nights of L-band flux in milliJanskys folded on the 12.5 hour rotation period of CU Vir (1.5 cycles displayed for clarity). Right and left circular polarisation are displayed in red/blue, respectively. Note that the main target of the MERLIN observation (the peak at phase ~0.8) was visible on all four nights. The secondary was also detected on ~1.5 nights. (b) The combined folded light curve for all four nights showing the complete 100% phase coverage. The two peaks were fitted with Gaussians to determine their separation and FHM size.

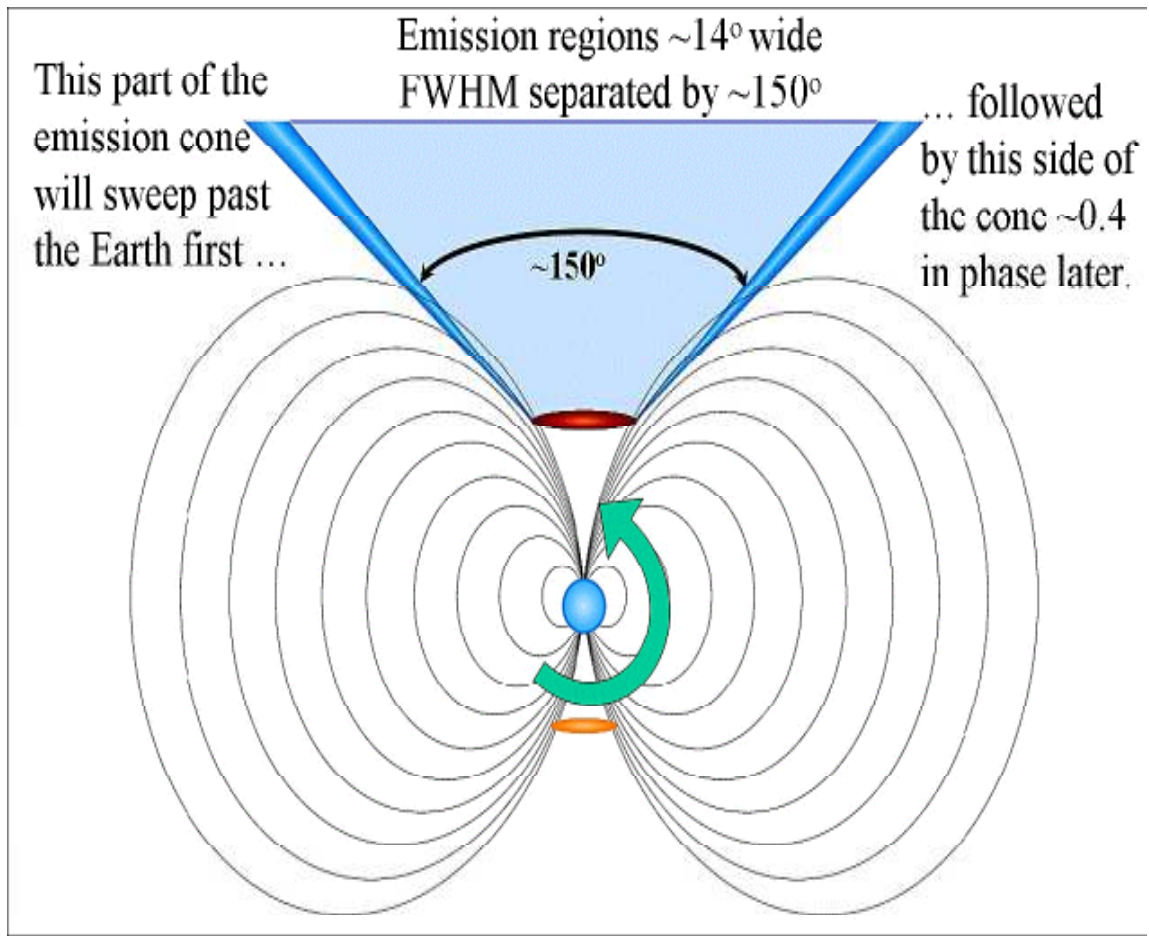

Figure 2: A schematic picture of CU Vir from a position above the rotation pole. As the star rotates anti-clockwise, the two sides of the emission cone from the stronger magnetic pole will sweep across the line-of-sight to Earth.